# Ferromagnetic Composite Self-Arrangement in Iron-Implanted Epitaxial Palladium Thin Films

A.I. Gumarov[1,2,a)], I.V. Yanilkin[1,2], R.V. Yusupov[2], A.G. Kiiamov[2],

L.R. Tagirov[1,2] and R.I. Khaibullin[1]

[1] Zavoisky Physical-Technical Institute, FRC Kazan Scientific Centre of RAS, 420029 Kazan, Russia

[2] Institute of Physics, Kazan Federal University, 420008 Kazan, Russia

[a)] Author to whom correspondence should be addressed: amir@gumarov.ru

**ABSTRACT**

We report on the formation of the dilute $Pd_{1-x}Fe_x$ compositions with tunable magnetic properties under an ion-beam implantation of epitaxial Pd thin films. Binary $Pd_{1-x}Fe_x$ alloys with a mean iron content $x$ of 0.025, 0.035 or 0.075 were obtained by the implantation of 40 keV $Fe^+$ ions into the palladium films on MgO (001) substrate to the doses of $0.5 \cdot 10^{16}$, $1.0 \cdot 10^{16}$ and $3.0 \cdot 10^{16}$ ions/cm$^2$, respectively. Structural and magnetic studies have shown that iron atoms occupy regular fcc-lattice Pd-sites without the formation of any secondary crystallographic phase. All the iron implanted Pd films reveal ferromagnetism at low temperatures (below 200 K) with both the Curie temperature and saturation magnetization determined by the implanted iron dose. In contrast to the magnetic properties of the molecular beam epitaxy grown $Pd_{1-x}Fe_x$ alloy films with the similar iron contents, the Fe-implanted Pd films possess weaker in-plane magnetocrystalline anisotropy, and, accordingly, a lower coercivity. The observed multiple ferromagnetic resonances in the implanted $Pd_{1-x}Fe_x$ films indicate a formation of a magnetically inhomogeneous state due to spinodal decomposition into regions, presumably layers, with identical crystal symmetry but different iron contents. The multiphase magnetic structure is robust with respect to the vacuum annealing at 770 K, though develops towards well-defined local Pd-Fe compositions.

# 1. Introduction

The renaissance of an interest to Pd$_{1-x}$Fe$_x$ alloys was triggered recently by their potential applications in superconducting spintronics, where diluted compositions with $x < 0.10$ serve as a weak ferromagnetic link in Josephson-junction structures for ultrahigh-speed cryogenic devices [1–6]. Traditionally, weak ferromagnetic Pd$_{1-x}$Fe$_x$ alloys have been fabricated using the magnetron sputtering [3–10] or molecular-beam epitaxy (MBE) [11,12], including our recent efforts focused on epitaxial thin films of Pd$_{1-x}$Fe$_x$ (0.01 < x < 0.1) on the MgO (001) substrate [13–16]. However, there is another technique with high potential to form dilute Fe-Pd alloys with tunable magnetic properties – the ion-beam implantation, widely used in modern microelectronics for silicon doping and microchip fabrication [17,18].

Although several reports on low-dose ion implantation of magnetic impurities into the pure-Pd films and foils can be found in the literature [19–22], they primarily report on the influence of irradiation-induced defects on the Curie temperature ($T_C$) and saturation magnetization ($M_s$). The majority of the reported implantation experiments had been done at liquid-helium temperature to avoid the annealing of radiation defects during ion bombardment. Also, the nature of these defects and the impact of thermal annealing on the magnetic properties were studied [19,20].

Targeted by the potential applications of Pd$_{1-x}$Fe$_x$ alloys in superconducting spintronics, we considered a high-dose (heavy) implantation of epitaxial palladium films with iron at room temperature (RT) to obtain weakly ferromagnetic layers with a minimum of irradiation-induced defects. Magnetometry and ferromagnetic resonance (FMR) studies have shown that Fe-implanted Pd films reveal ferromagnetism with the $T_C$ and $M_s$ strongly dependent on the implantation dose as it was expected. In contrast to the MBE-grown films with a similar iron content, the implanted films have lower in-plane magnetocrystalline anisotropy, and, accordingly, a lower coercive field. Depending on dose, revealed a presence of one to three magnetic phases with different $T_C$ and $M_s$. It looks that a pronounced separation into different Pd-Fe compositions takes place. These 'phases' are present already in the as-implanted samples and persist (though are modified) after the high-temperature vacuum annealing indicating the tendency towards separation into definite intrinsically-stable compositions. We discuss this finding in terms of spinodal decomposition of initially inhomogeneous distribution of Fe impurity into laminar composite state across the film thick-ness.

## 2. Experimental section

*2.1. Sample preparation and characterization*

Epitaxial Pd thin films on MgO (001) single-crystal substrate were used as the starting materials. The films were produced from the high purity Pd (99.98%) utilizing ultrahigh vacuum (UHV) molecular-beam epitaxy system (MBE, *by SPECS GmbH, Germany*). The three-step synthesis of epitaxial Pd films is described in detail in Ref. [14]. The $^{56}$Fe$^+$ ion implantation was performed with the *ILU*-3 accelerator at fixed ion energy of 40 keV and ion currents of 2-3 $\mu$A as measured at the sample. To obtain Pd-Fe alloys with various iron concentrations, Fe$^+$ ion doses of $0.5 \cdot 10^{16}$, $1.0 \cdot 10^{16}$ and $3.0 \cdot 10^{16}$ ions/cm$^2$ (*sample labels* S0_5, S1_0 and S3_0, *respectively*) were achieved varying the irradiation time. An expected depth profile of Fe-concentration was calculated in advance for $1.0 \cdot 10^{16}$ ions/cm$^2$ dose following the SRIM-2013 (TRIM) algorithm [23], and the Fe$^+$ doses were adjusted accordingly to yield the samples with the mean iron contents of $\bar{c}_{Fe} \approx 2.5$, 3.5 and 7.5 at.%. Note that the original TRIM approach does not account for the sputtering of the target front surface, changes in target composition and diffusion during the implantation. In order to obtain comparable thicknesses of the resulting alloy layer, initial Pd films had the thicknesses of 40 nm, 60 nm and 80 nm for the three above-mentioned concentrations of iron, respectively. Film thickness was measured prior to and after Fe$^+$ implantation with the *Bruker DektakXT* stylus profilometer with an accuracy of ±0.5 nm. The implanted samples had been cut into parts, and one of them was post-annealed in vacuum at 770 K for 20 min.

The formation of the Pd$_{1-x}$Fe$_x$ alloy with cubic symmetry corresponding to the crystal lattice of pristine Pd matrix was justified *ex situ* with X-ray diffraction (XRD, *Bruker D8 Advance*). XRD studies were performed utilizing the Cu-K$_\alpha$ radiation ($\lambda = 0.15418$ nm) in the Bragg–Brentano geometry, in the range of $2\theta$ angles of 40 to 72 degrees. XRD patterns are presented in Fig. 1. No sign of any secondary phase (*e.g.* metallic Fe nanoparticles) could be found at the experimental sensitivity level. The $\theta$-$2\theta$ scan (Fig. 1a) shows notable shift and broadening of the S3_0 implanted film (002) maximum relative to the pure Pd and the MBE-grown Pd$_{1-x}$Fe$_x$ films due to the lattice parameter modification (see detailed analysis in Ref. [14]) and inhomogeneous distribution of the iron across the film thickness, respectively. XRD $\varphi$-scan for S3_0 sample is shown in Fig. 1b in comparison with that for the MBE-grown Pd$_{0.92}$Fe$_{0.08}$ film (Fig. 1c). XRD data clearly indicate the cubic symmetry, cube-on-cube epitaxy and single-crystalline structure of the implanted film.

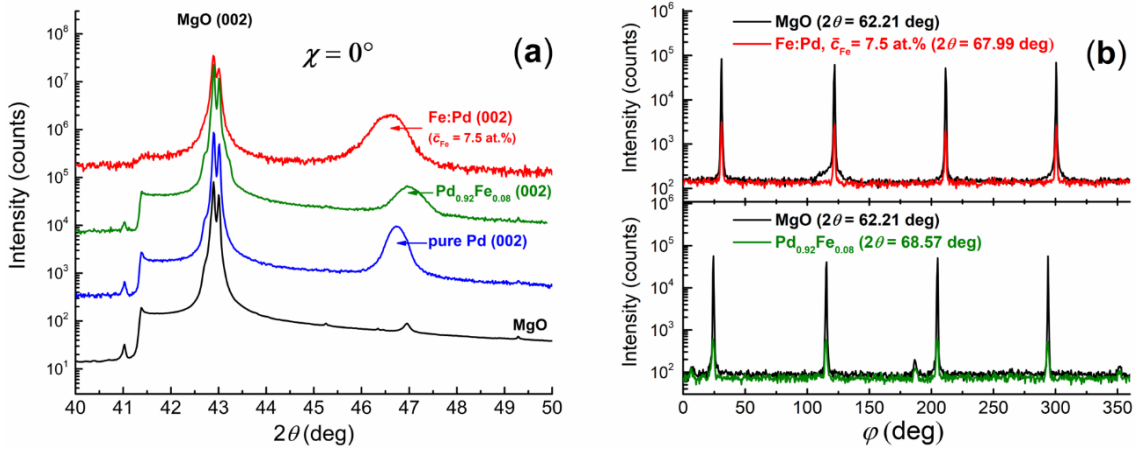

**FIG. 1.** X-ray diffraction patterns: $2\theta$-scans (a) and $\varphi$-scans (b,c) for the S3_0 implanted film (red line), MBE-grown epitaxial $Pd_{0.92}Fe_{0.08}$ film (green line); pure 40 nm Pd film (blue line) and MgO substrate (black line). Eulerian cradle angle $\chi$ was set to 0 degree in $2\theta$-scans and was equal to 45 degrees in $\varphi$-scans in order to detect <202> XRD-maxima (angle $\varphi$ is arbitrary).

The iron concentration profiles were obtained by X-ray photoelectron spectroscopy (XPS) in combination with depth profiling by the $Ar^+$-ion etching technique. Overview XPS spectrum (Fig. 2a) for the implanted film demonstrates the set of lines corresponding only to iron and palladium elements. No alien elements were detected within the sensitivity of our spectrometer (~ 0.1%). Figures 2b and 2c show high-resolution spectra of core-shell electrons $Fe_{2p}$ and $Pd_{3d}$ recorded with the 0.1 eV step and the analyzer transmission energy of $E = 25$ eV. Binding energies and line shapes are characteristic of Pd-Fe alloy studied earlier by XPS [14]. For comparison, we recorded also high resolution XPS-spectra of the pure iron and palladium films. Annealing did not lead to any significant modification of the shape and position of the XPS-lines. The obtained results signify that Pd-Fe alloys synthetized by ion implantation or molecular-beam epitaxy [14] exhibit the same dissolving mechanism: iron atoms substitute for the palladium ones in the lattice without formation of iron nanoparticles. The last is confirmed by the absence of a characteristic peak for metallic iron ($Fe^0$) at the binding energy of 706.7 eV [24] (within the sensitivity of our XPS setup).

Depth profiling of the samples was carried out in a step-by-step manner, combining etching of the sample with argon ions and recording the high-resolution XPS-spectra after the each etching step. The $Fe_{2p}$ and $Pd_{3d}$ spectra were used to calculate the concentration of iron impurities. The iron impurity was detected across the entire thickness of every $Fe^+$-irradiated palladium film. The film thickness after the iron implantation according to the profilometry data was equal to 36 nm, 50 nm and 63 nm for doses of $0.5 \cdot 10^{16}$, $1.0 \cdot 10^{16}$, и $3.0 \cdot 10^{16}$ ions/cm$^2$, respectively. Thus, a significant sputtering of the front surface of the films during the implantation occurred.

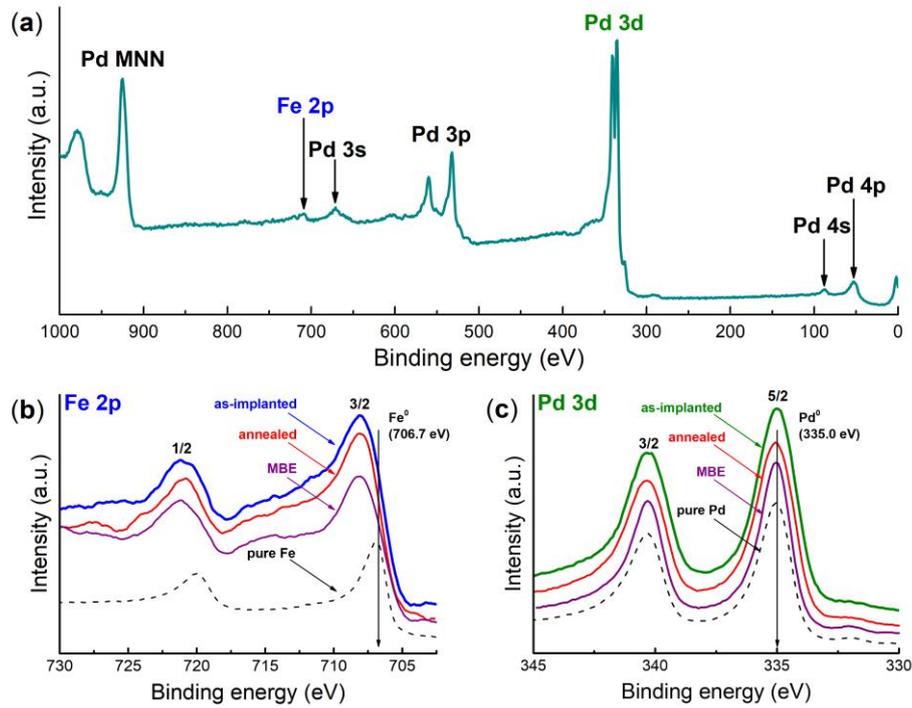

**FIG. 2.** a) A survey XPS spectrum recorded from the S3_0 sample; b) and c) are the high-resolution XPS spectra of the inner shell Fe$_{2p}$ and Pd$_{3d}$ electrons, respectively, recorded for the as-implanted, annealed in vacuum iron-implanted Pd films, Pd$_{0.92}$Fe$_{0.08}$ alloy synthesized by MBE and reference films of pure iron and palladium grown separately.

Distribution profiles of the iron impurity for the three doses are shown in Fig. 3. The shape of the distribution profile changes with the implantation dose and after the subsequent thermal annealing. Obtained distributions have essentially a stepped shape. Partial redistribution of the iron impurity over the film thickness takes place during the annealing, especially notable for the S0_5 and S3_0 samples. The most of the impurity is located up to a depth of 35 nm from the surface for the S0_5 and S1_0 samples, and to a depth of 55 nm for the S3_0 sample. A local maximum of the impurity concentration is observed near the film-substrate interface of the as-implanted S3_0 sample. The iron impurity has not been found inside the substrate for any implantation dose used in this work. Thus, the MgO substrate serves a barrier for the implanted iron impurity.

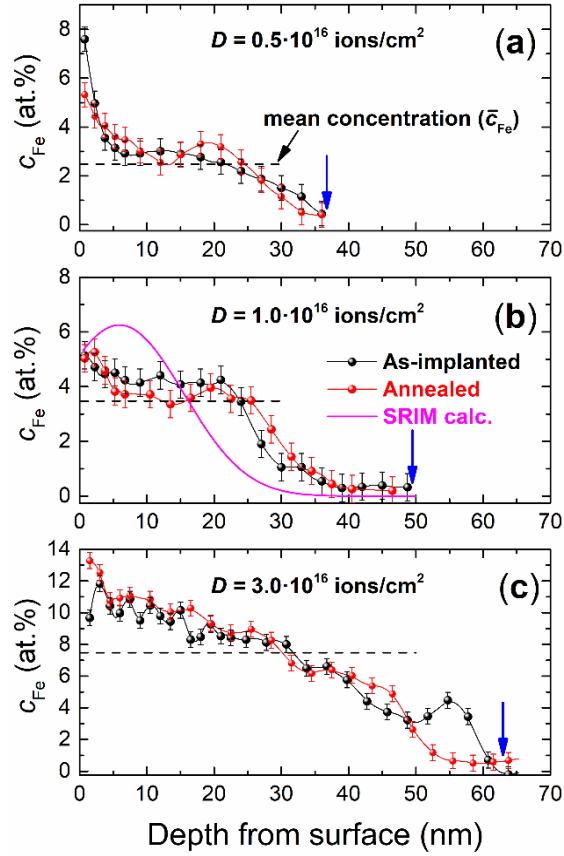

**FIG. 3.** Distribution profiles of the iron impurity in S0_5 (a), S1_0 (b) and S3_0 (c) samples, obtained both before (black symbols and curves) and after the thermal annealing (red symbols and curves). The pink solid line in panel (b) shows the calculated profile of the iron distribution in the palladium matrix (see text). Arrows indicate the position of the film/substrate interface.

*2.2. Magnetic properties*

Magnetic properties of the iron-implanted Pd films were measured utilizing the vibrating sample magnetometry (VSM, *Quantum Design PPMS-9*) and ferromagnetic resonance (FMR, *Bruker ESP300*) spectroscopy in the temperature range of 5 – 300 K with the magnetic field applied either *in-plane* or *out-of-plane* of the films. To determine $M_s$ of the synthesized Pd-Fe alloy films, the combined diamagnetic contribution of the MgO substrate and paramagnetic one from the impurities in it were subtracted from the raw VSM data. Then, the saturation magnetic moment was recalculated to the number of Bohr magnetons ($\mu_B$) per implanted $Fe^+$-ion.

Figure 4a shows magnetic hysteresis loops for palladium films implanted with three different doses of iron. For the S0_5 sample, a rather low specific saturation magnetic moment of ≈ 3 $\mu_B$/Fe was obtained, which is not far from 3.7 $\mu_B$/Fe reported in [8] for the magnetron sputtered film of $Pd_{0.99}Fe_{0.01}$. On the other hand, the specific moment of ≈ 6.5 $\mu_B$/Fe was found for the MBE grown epitaxial film of 20 nm thickness with $x = 0.019$ [16], and ≈ 6.9 $\mu_B$/Fe – for the bulk samples [25–27]. Certainly, the higher value of the magnetic moment per iron correlates with the synthesis techniques providing more uniform distribution of Fe (bulk and MBE samples), while that providing less uniform distribution (magnetron sputtering and ion implantation) lead to its reduction.

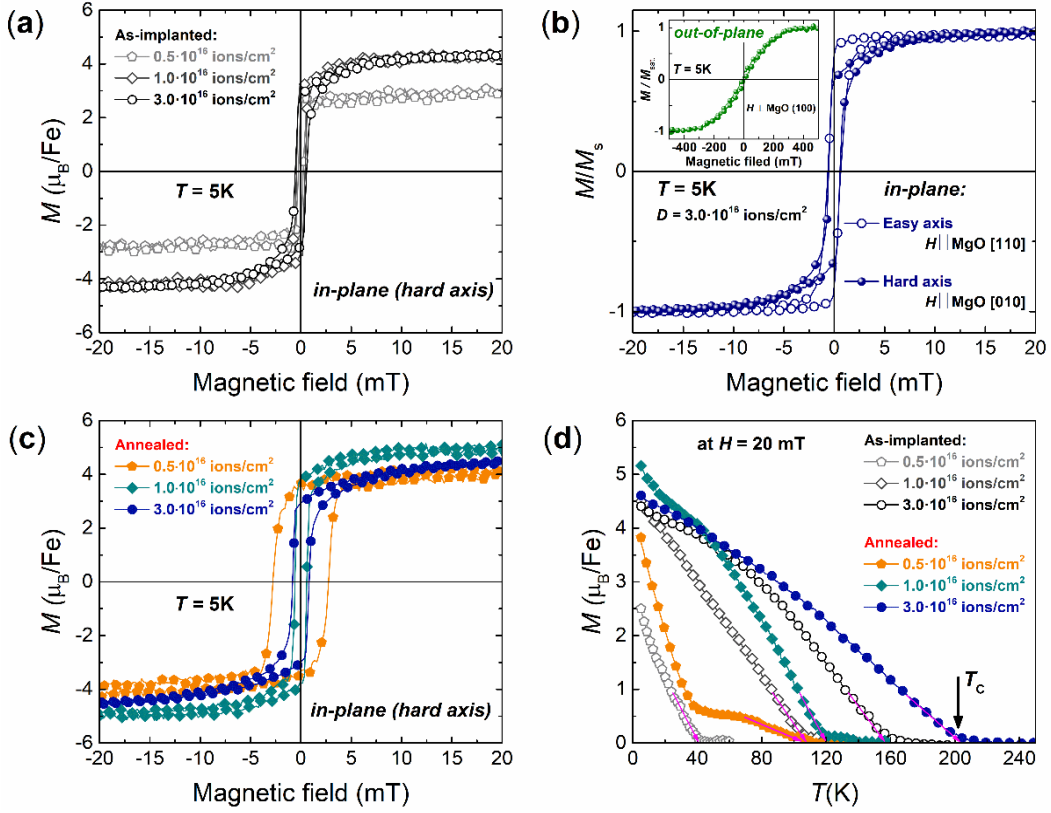

**FIG. 4.** Magnetic hysteresis loops measured with the magnetic field applied along the hard in-plane magnetization axis for the S0_5, S1_0 and S3_0 samples (a); along the in-plane easy and hard magnetization axes (b); as well as out-of-plane (inset to panel (b)), and after the thermal annealing of the samples (c). Temperature dependences of the saturation magnetization both before (open symbols) and after (solid symbols) annealing (d).

Figure 4b presents the magnetic hysteresis loops measured with three orientations of the applied magnetic field $H$ with respect to the S3_0 sample plane: the *in-plane* with $H//[110]$ and $H//[010]$ crystallographic directions of the MgO (001) substrate and the *out-of-plane* geometry with $H\|[001]$ (Fig. 4b, inset). The rectangular shape of the loop shows that the <110> directions are the easy *in-plane* directions, while the <010> are the hard *in-plane* ones in full agreement with the properties of the MBE-grown epitaxial Pd$_{1-x}$Fe$_x$ films [13–16]. Reversible character and a higher value of the saturating magnetic field in the *out-of-plane* geometry indicate that the obtained material is the easy-plane magnetic system. Qualitatively similar results were obtained for the S0_5 and S1_0 samples.

Figure 4c shows the magnetic hysteresis loops for the implanted palladium films of Fig. 4a measured after their thermal annealing. The annealing leads to an increase in the magnetization and coercive field of the implanted samples. Nevertheless, both the as-implanted and post-annealed Pd films reveal weaker in-plane magnetocrystalline anisotropy and a lower coercivity in comparison with the MBE grown $Pd_{1-x}Fe_x$ films with similar iron concentration (see Fig. S1 of Supplementary material). At the same time, from the temperature dependences of the magnetic moments, Fig. 4d, it follows that $T_C$ for all iron implanted Pd films increases after high-temperature annealing in vacuum. The magnetometry results for as-implanted and post-annealed Pd films are summarized in Table 1.

**Table 1.** Experimentally evaluated magnetic properties of as-prepared and post-annealed thin palladium films implanted with Fe-ions to different doses $D$.

| Sample label | | $D$, $10^{16}$ ions/cm$^2$ | $\bar{c}_{Fe}$, at.% | $M_s$, $\mu_B$/Fe | $B_c^*$, mT | $T_C$, K |
|---|---|---|---|---|---|---|
| As-implanted | S0_5 | 0.5 | 2.5 | 3.0 | 0.24 | 40 |
| | S1_0 | 1.0 | 3.5 | 4.5 | 0.43 | 107 |
| | S3_0 | 3.0 | 7.5 | 4.4 | 0.56 | 155 |
| Post-annealed | S0_5a | 0.5 | 2.5 | 4.2 | 2.76 | 105 |
| | S1_0a | 1.0 | 3.5 | 5.0 | 0.58 | 120 |
| | S3_0a | 3.0 | 7.5 | 4.6 | 0.83 | 203 |

*The coercivity $B_c$ is measured at 5 K

The most unusual results for the Fe-implanted thin palladium films were obtained with the ferromagnetic resonance (FMR) spectroscopy [28]. Temperature evolution of the FMR spectrum with the magnetic field applied perpendicular to the as-implanted S3_0 sample plane is presented in Fig. 5a. A consecutive appearance of the resonances is observed as if ferromagnetic phases emerge sequentially with the temperature decrease. FMR spectra of the same sample for the magnetic field applied along the [110] (in-plane) and [001] (out-of-plane) directions are shown in Fig. 5b (black lines). The modification of the FMR spectrum of the S3_0 sample after the thermal annealing is illustrated also by Fig. 5b (sample S3_0a, red lines). In-plane resonance lines shift after the annealing towards the lower fields, and the two out-of-plane lines – towards the higher fields being a common feature of the spectra evolution for all the samples. The multiphase response of the magnetic system persists after the annealing.

The angular behavior of the resonance fields of both the as-implanted and the annealed samples is typical for the easy-plane systems: in-plane resonances reside in lower magnetic fields with respect to the electron paramagnetic resonance lines (a comb centered around 330 mT originating from the 3d-ion impurities in the substrate), while the out-of-plane resonances are found at higher fields due to the thin-film 2D shape anisotropy [28]. For comparison, the FMR spectra of the homogeneous epitaxial $Pd_{0.92}Fe_{0.08}$ film [13] are presented in Fig. 5b.

Resonance fields for the annealed Fe-implanted Pd-film, sample S3_0a, in the out-of-plane orientation have the same values as that for the epitaxial $Pd_{1-x}Fe_x$ films [16] with the iron concentrations of $x = 0.025$, $x = 0.04$ and $x = 0.063$. The S0_5a sample with the lowest dose reveals only one FMR signal at $T \leq 40$ K (inset to Fig. 5b and Fig. S2a of the Supplementary material), and the S1_0a sample – two FMR signals (see Fig. S2b).

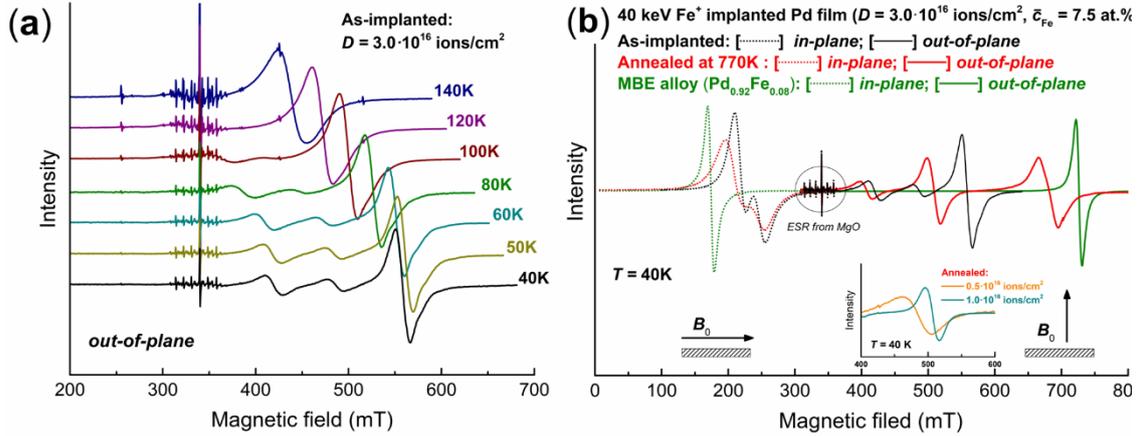

**FIG. 5.** FMR spectra of the epitaxial Pd film implanted with iron to the dose of $3.0 \cdot 10^{16}$ ions/cm$^2$: (a) temperature evolution for the as-implanted S3_0 sample in the *out-of-plane* geometry; (b) *in-plane* and *out-of-plane* spectra of the as-implanted (sample S3_0, black curves) and the annealed (sample S3_0a, red curves) films, as well as the spectra of the MBE-grown epitaxial $Pd_{0.92}Fe_{0.08}$ film (green curves). The inset shows the resonance lines of the annealed S0_5a and S1_0a samples ($\nu = 9.416$ GHz, $T = 40$ K).

### 3. Discussion

Observation of multiple ferromagnetic resonances unambiguously indicates an occurrence of the corresponding set of rather well-defined, in the sense of the $M_s$ values, ferromagnetic components in our Fe-implanted palladium films. This, in turn, is a manifestation of the Pd-Fe system separation into definite compositions. In our opinion, this can take place due to a kind of a spinodal decomposition [29–32] occurring under non-equilibrium conditions in the course of the implantation, that lead, most probably, to a formation of a laminar composite magnetic structure in the implanted $Pd_{1-x}Fe_x$ films.

Indeed, because of the doping technique, the implant distribution is inhomogeneous *a priori*, see Fig. 3. The distribution shape changes with increasing the dose because of the front surface sputtering during the irradiation, and diffusion stimulated by the annealing [33] due to a retardation of the penetrating ions with the kinetic energy mostly transferred to the heat. This could be a reason of an absence of significant evolution of the depth distribution profiles and magnetic hysteresis loops, Fig. 4, upon post-implantation thermal annealing. Single-crystal state of the Fe-implanted Pd films (see Fig. 1) also points at a dynamic annealing of radiation defects in the films directly during the ion irradiation. The S0_5 sample implanted with the lowest dose of $0.5 \cdot 10^{16}$ ions/cm$^2$ (the lowest time of implantation) seems to acquire minimal self-annealing, that is why it shows maximal changes in the hysteresis loop shape, $M_s$ and $T_C$ upon *ex situ* annealing, Fig. 4, c and d.

In spite of inhomogeneous distribution of iron atoms, there are no indications of sequential reversals of different magnetic phases in hysteresis loops (Figs. 4, a to c). Indeed, if the remagnetization mechanism is a movement of domain walls, then the most "soft" region (sublayer) of the film triggers at low fields and drags the coupled neighboring sublayers to complete reversal of the entire magnetic moment of a film. No multiphase, laminar nature of magnetism could be recognized in the quasi-static magnetic moment measurements, Fig. 4 and Fig. S1 of the Supplementary material. The regions with high iron content could lead to an increase in coercivity due to the domain-wall pinning by the coupling with them (see Fig. S1).

Our FMR studies revealed that in the spectrum for a high-dose S3_0 sample the number of resonance lines increases from one to three with decreasing the temperature (Fig. 5a). This feature is quite robust, because the high-temperature annealing of the sample does not qualitatively change its FMR response (Fig. 5b). Probably, the inhomogeneous profile of the iron distribution in the implanted palladium films leads to the formation of a laminar multiphase magnetic system with different temperatures of ferromagnetic ordering and saturation magnetizations for each individual phase (layer). In spite of a direct contact of these phases (layers), and the resulting coupling between them, they show robust individual responses with the angular behavior exactly mimicking the FMR response of thin ferromagnetic film with the easy-plane shape anisotropy [13,16,28].

Moreover, for palladium films implanted with a minimum ($0.5 \cdot 10^{16}$ ions/cm$^2$, S0_5) and medium ($1.0 \cdot 10^{16}$ ions/cm$^2$, S1_0) iron doses, after bringing them to thermodynamic equilibrium by the annealing, we observed one FMR line for the first, and two FMR lines for the second (Fig. S2 in the Supplementary material). The closeness of the resonance fields for the selected lines (see Fig. S3 of the Supplementary material) supports a formulation of a simple model of spinodal decomposition into isostructural laminar structure [29], each layer possessing a certain, quite well-defined concentration of iron in the palladium. That is, the system evolves from a metastable non-equilibrium state created by ion implantation, to its energetically favorable state by the formation of several spatially-extended (macroscopic) "*stable phases*". The shift of the resonance line at ~ 420 mT in Fig. 5b towards lower resonance fields, as well as the appearance of inflections in the temperature dependence of magnetization for low and medium doses (Fig. 4d) after thermal annealing support this conjecture. As an ultimate hypothesis on the microscopic scale, the FMR technique reveals fine preferences for certain compositions of Pd-Fe system that highlights the effects of preferential local order.

Disclosure of this previously unknown tendency of the $Pd_{1-x}Fe_x$ binary system in the palladium-rich ($x < 0.10$) composition range opens up new prospects for the use of these materials in superconducting spintronic heterostructures: the reported intrinsically stable compositions of $x = 0.025$, $x = 0.04$, $x = 0.063$ (preliminary estimates for uncoupled layers) should be utilized ensuring the magnetic homogeneity of the ferromagnetic layers. Ion-beam implantation of epitaxial palladium films with relatively low doses of iron can serve a route for a synthesis of near-homogeneous low-temperature ferromagnets.

## 4. Conclusions

To summarize, we synthesized Pd-Fe alloys with a mean iron content in the range of 2.5-7.5 at.% using high-dose Fe-ion implantation to epitaxial Pd-films at RT. Iron-implanted Pd films retain fcc cubic crystal structure typical for binary Pd-Fe alloys with low content of iron. All synthesized samples exhibit ferromagnetism at low temperatures. The Curie temperature and saturation magnetization depend on the dose of implantation and are modified under subsequent thermal annealing. Magnetic hysteresis loops for the implanted palladium films are narrower than those for Pd-Fe alloys with comparable concentrations grown by MBE. FMR and VSM studies indicate that implantation of epitaxial palladium films with iron leads to the unusual formation of a laminar magnetic composite in otherwise crystallographically homogeneous film. The multiphase behavior is discussed within the hypothesis of spinodal decomposition of the initially non-equilibrium distribution of inhomogeneously delivered dopant to isostructural, laterally extended laminar sub-structures.

**CRediT authorship contribution statement:**

**Gumarov A.I.:** Conceptualization, Sample preparation, Measurements, Data curation, Writing – original draft; **Yanilkin I.V.:** Measurements, Data curation; **Yusupov R.V.**: Investigation, Writing – review & editing; **Kiiamov A.G.:** X-ray data acquisition & curation; **Tagirov L.R.:** Formal analysis, Writing – original draft, Writing – review & editing; **Khaibullin R.I.:** Conceptualization, Methodology, Sample preparation, Supervision, Funding acquisition.

**Declaration of competing interest:**

The authors declare that they have no competing interests.

**Acknowledgements:**

Authors thank V.I. Nuzhdin and V.F. Valeev for technical assistance with ion implantation experiments.

**Funding:**

This work was supported by the RFBR Grant No. 20-02-00981.

**Data available on request from the authors:**

The data that support the findings of this study are available from the corresponding author upon a reasonable request.

**References**

[1] I.A. Golovchanskiy, V. V. Bolginov, N.N. Abramov, V.S. Stolyarov, A. Ben Hamida, V.I. Chichkov, D. Roditchev, V.V. Ryazanov, Magnetization dynamics in dilute Pd1-xFex thin films and patterned microstructures considered for superconducting electronics, J. Appl. Phys. 120 (2016) 163902. https://doi.org/10.1063/1.4965991.

[2] V.V. Bol'ginov, V.S. Stolyarov, D.S. Sobanin, A.L. Karpovich, V.V. Ryazanov, Magnetic switches based on Nb-PdFe-Nb Josephson junctions with a magnetically soft ferromagnetic interlayer, JETP Lett. 95 (2012) 366–371. https://doi.org/10.1134/S0021364012070028.

[3] T.I. Larkin, V.V. Bol'ginov, V.S. Stolyarov, V.V. Ryazanov, I.V. Vernik, S.K. Tolpygo, O.A. Mukhanov, Ferromagnetic Josephson switching device with high characteristic voltage, Appl. Phys. Lett. 100 (2012) 222601. https://doi.org/10.1063/1.4723576.

[4] V.V. Ryazanov, V.V. Bol'ginov, D.S. Sobanin, I.V. Vernik, S.K. Tolpygo, A.M. Kadin, O.A. Mukhanov, Magnetic Josephson Junction Technology for Digital and Memory Applications, Phys. Procedia. 36 (2012) 35–41. https://doi.org/10.1016/J.PHPRO.2012.06.126.

[5] I.V. Vernik, V.V. Bol'Ginov, S.V. Bakurskiy, A.A. Golubov, M.Y. Kupriyanov, V.V. Ryazanov, O.A. Mukhanov, Magnetic josephson junctions with superconducting interlayer for cryogenic memory, IEEE Trans. Appl. Supercond. 23 (2013) 1701208–1701208. https://doi.org/10.1109/TASC.2012.2233270.

[6] J.A. Glick, R. Loloee, W.P. Pratt,, N.O. Birge, Critical Current Oscillations of Josephson Junctions Containing PdFe Nanomagnets, IEEE Trans. Appl. Supercond. 27 (2016) 1–5. https://doi.org/10.1109/TASC.2016.2630024.

[7] H.Z. Arham, T.S. Khaire, R. Loloee, W.P. Pratt, N.O. Birge, Measurement of spin memory lengths in PdNi and PdFe ferromagnetic alloys, Phys. Rev. B - Condens. Matter Mater. Phys. 80 (2009) 174515. https://doi.org/10.1103/PhysRevB.80.174515.

[8] L.S. Uspenskaya, A.L. Rakhmanov, L.A. Dorosinskii, S.I. Bozhko, V.S. Stolyarov, V.V. Bolginov, Magnetism of ultrathin Pd 99 Fe 01 films grown on niobium, Mater. Res. Express. 1 (2014) 036104. https://doi.org/10.1088/2053-1591/1/3/036104.

[9] L.S. Uspenskaya, I.V. Shashkov, Influence of $Pd_{0.99}Fe_{0.01}$ film thickness on magnetic properties, Phys. B Condens. Matter. (2017). https://doi.org/10.1016/j.physb.2017.09.089.

[10] V.V. Bol'ginov, O.A. Tikhomirov, L.S. Uspenskaya, Two-component magnetization in $Pd_{99}Fe_{01}$ thin films, JETP Lett. 105 (2017) 169–173. https://doi.org/10.1134/S0021364017030055.

[11] M. Ewerlin, B. Pfau, C.M. Günther, S. Schaffert, S. Eisebitt, R. Abrudan, H. Zabel, Exploration of magnetic fluctuations in PdFe films, J. Phys. Condens. Matter. 25 (2013) 266001. https://doi.org/10.1088/0953-8984/25/26/266001.

[12] I.A. Garifullin, D.A. Tikhonov, N.N. Garif'yanov, M.Z. Fattakhov, K. Theis-Bröhl,

K. Westerholt, H. Zabel, Possible reconstruction of the ferromagnetic state under the influence of superconductivity in epitaxial V/Pd$_{1-x}$Fe$_x$ bilayers, Appl. Magn. Reson. 22 (2002) 439–452. https://doi.org/10.1007/BF03166124.

[13]  A. Esmaeili, I.R. Vakhitov, I.V. Yanilkin, A.I. Gumarov, B.M. Khaliulin, B.F. Gabbasov, M.N. Aliyev, R.V. Yusupov, L.R. Tagirov, FMR Studies of Ultra-Thin Epitaxial Pd$_{0.92}$Fe$_{0.08}$ Film, Appl. Magn. Reson. 49 (2018) 175–183. https://doi.org/10.1007/s00723-017-0946-1.

[14]  A. Esmaeili, I. V. Yanilkin, A.I. Gumarov, I.R. Vakhitov, B.F. Gabbasov, A.G. Kiiamov, A.M. Rogov, Y.N. Osin, A.E. Denisov, R.V. Yusupov, L.R. Tagirov, Epitaxial growth of Pd$_{1-x}$Fe$_x$ films on MgO single-crystal substrate, Thin Solid Films. 669 (2019) 338–344. https://doi.org/10.1016/j.tsf.2018.11.015.

[15]  W.M. Mohammed, I.V. Yanilkin, A.I. Gumarov, A.G. Kiiamov, R.V. Yusupov, L.R. Tagirov, Epitaxial growth and superconducting properties of thin-film PdFe/VN and VN/PdFe bilayers on MgO(001) substrates, Beilstein J. Nanotechnol. 11 (2020) 807–813. https://doi.org/10.3762/bjnano.11.65.

[16]  A. Esmaeili, I.V. Yanilkin, A.I. Gumarov, I.R. Vakhitov, B.F. Gabbasov, R.V. Yusupov, D.A. Tatarsky, L.R. Tagirov, Epitaxial thin-film Pd1-xFex alloy: a tunable ferromagnet for superconducting spintronics, Sci. China Mater. (2020). https://doi.org/10.1007/s40843-020-1479-0.

[17]  L.A. Larson, J.M. Williams, M.I. Current, Ion implantation for semiconductor doping and materials modification, Rev. Accel. Sci. Technol. Accel. Appl. Ind. Environ. 04 (2012) 11–40. https://doi.org/10.1142/S1793626811000616.

[18]  R.W. Hamm, M.E. Hamm (eds.), Industrial accelerators and their applications, World Scientific, 2012. https://doi.org/10.1142/7745.

[19]  J.D. Meyer, B. Stritzker, Reduced Susceptibility of Irradiated Palladium, Phys. Rev. Lett. 48 (1982) 502–505. https://doi.org/10.1103/PhysRevLett.48.502.

[20]  M. Hitzfeld, P. Ziemann, W. Buckel, H. Claus, Ferromagnetism of alloys: A new probe in ion implantation, Solid State Commun. 47 (1983) 541–544. https://doi.org/10.1016/0038-1098(83)90495-7.

[21]  M. Hitzfeld, P. Ziemann, W. Buckel, H. Claus, Ferromagnetism of Pd-Fe alloys produced by low-temperature ion implantation, Phys. Rev. B. 29 (1984) 5023–5030. https://doi.org/10.1103/PhysRevB.29.5023.

[22]  H. Claus, N.C. Koon, Effect of oxygen on the electronic properties of Pd, Solid State Commun. 60 (1986) 481–484. https://doi.org/10.1016/0038-1098(86)90721-0.

[23]  J.F. Ziegler, M.D. Ziegler, J.P. Biersack, SRIM - The stopping and range of ions in matter (2010), Nucl. Instruments Methods Phys. Res. Sect. B Beam Interact. with Mater. Atoms. 268 (2010) 1818–1823. https://doi.org/10.1016/j.nimb.2010.02.091.

[24]  M.C. Biesinger, B.P. Payne, A.P. Grosvenor, L.W.M. Lau, A.R. Gerson, R.S.C. Smart,

Resolving surface chemical states in XPS analysis of first row transition metals, oxides and hydroxides: Cr, Mn, Fe, Co and Ni, Appl. Surf. Sci. 257 (2011) 2717–2730. https://doi.org/10.1016/j.apsusc.2010.10.051.

[25] J. Crangle, W.R. Scott, Dilute ferromagnetic alloys, J. Appl. Phys. 36 (1965) 921–928. https://doi.org/10.1063/1.1714264.

[26] G.J. Nieuwenhuys, Magnetic behaviour of cobalt, iron and manganese dissolved in palladium, Adv. Phys. 24 (1975) 515–591. https://doi.org/10.1080/00018737500101461.

[27] B. Heller, K.H. Speidel, R. Ernst, A. Gohla, U. Grabowy, V. Roth, G. Jakob, F. Hagelberg, J. Gerber, S.N. Mishra, P.N. Tandon, Transient field measurement in the giant moment PdFe alloy, Nucl. Instruments Methods Phys. Res. Sect. B Beam Interact. with Mater. Atoms. 142 (1998) 133–138. https://doi.org/10.1016/S0168-583X(98)00260-2.

[28] M. Farle, Ferromagnetic resonance of ultrathin metallic layers, Reports Prog. Phys. 61 (1998) 755–826. https://doi.org/10.1088/0034-4885/61/7/001.

[29] J.W. Cahn, On spinodal decomposition, Acta Metall. 9 (1961) 795–801. https://doi.org/10.1016/0001-6160(61)90182-1.

[30] V.P. Skripov, A.V. Skripov, Spinodal decomposition (phase transitions via unstable states), Sov. Phys. - Uspekhi. 22 (1979) 389–410. https://doi.org/10.1070/PU1979v022n06ABEH005571.

[31] Systems Far from Equilibrium (ed. by Luis Garrido), Springer-Verlag, Berlin-Heidelberg-New York, 1980. https://doi.org/ 10.1007/BFb0025609.

[32] Phase Transformations in Materials (ed. by Gernot Kostorz), Wiley VCH Verlag GmbH, Weinheim, 2001. https://doi.org/10.1002/352760264X.

[33] A.A. Achkeev, R.I. Khaibullin, L.R. Tagirov, A. Mackova, V. Hnatowicz, N. Cherkashin, Specific features of depth distribution profiles of implanted cobalt ions in rutile $TiO_2$, Phys. Solid State. 53 (2011) 543–553. https://doi.org/10.1134/S1063783411030024.

# Ferromagnetic Composite Self-Arrangement in Iron-Implanted Epitaxial Palladium Thin Films

A.I. Gumarov[1,2,a)], I.V. Yanilkin[1,2], R.V. Yusupov[1,2], A.G. Kiiamov[2],

L.R. Tagirov[1,2] and R.I. Khaibullin[1]

[1] Zavoisky Physical-Technical Institute, FRC Kazan Scientific Centre of RAS, 420029 Kazan, Russia

[2] Institute of Physics, Kazan Federal University, 420008 Kazan, Russia

[a)] Author to whom correspondence should be addressed: amir@gumarov.ru

## Comparison of $Pd_{1-x}Fe_x$ films data: Ion implantation *vs*. MBE

**VSM studies**

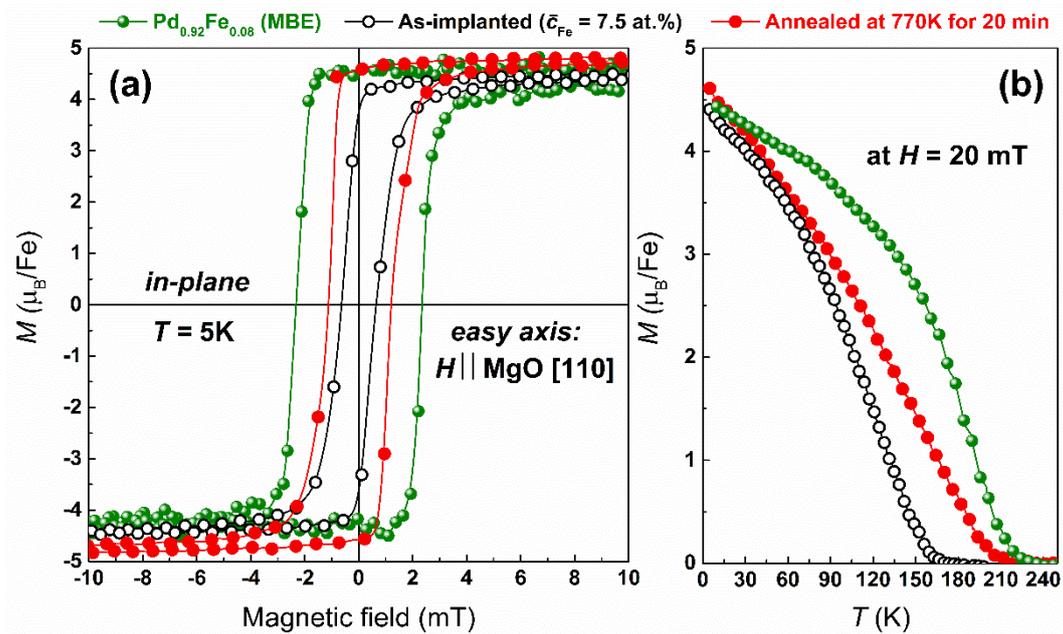

**FIG S1.** Hysteresis loops at 5 K (a) and $M(T)$ curves (b) measured for 80 nm epitaxial Pd film implanted with 40 keV $Fe^+$-ions with the dose of $3 \cdot 10^{16}$ ions/cm$^2$ (○); the same sample post-annealed in vacuum at 770 K for 20 min (●); epitaxial film of $Pd_{0.92}Fe_{0.08}$ alloy deposited by MBE to (100) MgO substrate (●).

**FMR studies**

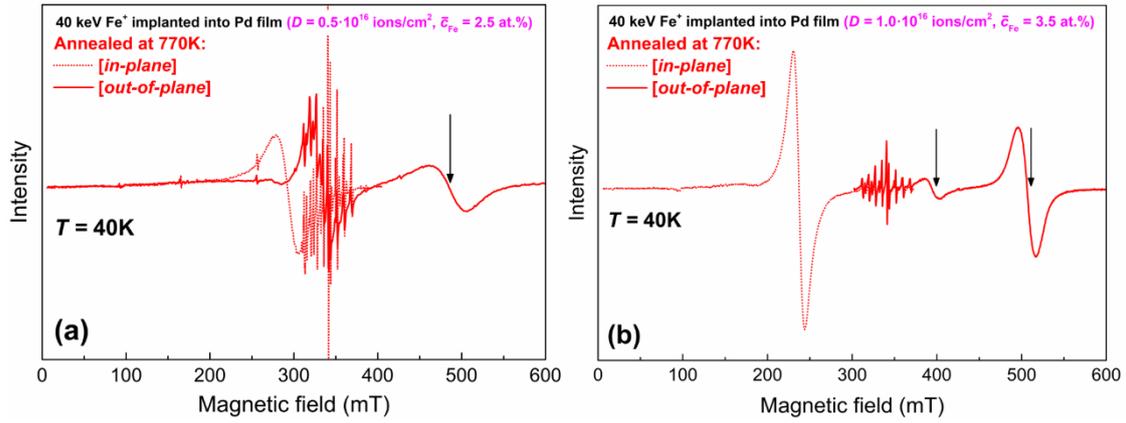

**FIG S2.** FMR spectra obtained in the in-plane (dotted line) and out-of-plane (solid line) measurement geometries for epitaxial Pd thin film implanted with 40 keV $Fe^+$-ions with the dose of $0.5 \cdot 10^{16}$ ions/cm$^2$ (a) and $1.0 \cdot 10^{16}$ ions/cm$^2$ (b). The two samples have been measured after vacuum annealing at 770 K for 20 min. Arrows indicate the observed resonance lines in the out-of-plane orientation.

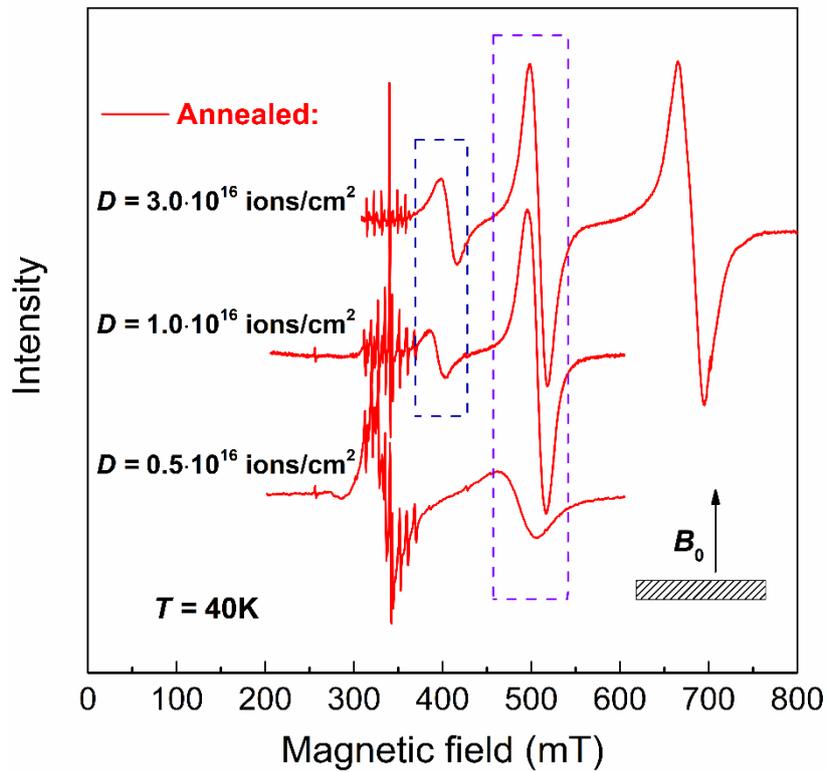

**FIG S3.** FMR spectra for all three Fe-implanted palladium films after the annealing for comparison of the resonance fields.